# Classification Modeling with RNN-based, Random Forest, and XGBoost for Imbalanced Data: A Case of Early Crash Detection in ASEAN-5 Stock Markets

**Deri Siswara[1*], Agus M. Soleh[2], Aji Hamim Wigena[3]**

[1,2,3] Department of Statistics and Data Sciences, Faculty of Mathematics and Natural Sciences, IPB University, Indonesia

**Abstract.**

**Purpose:** This research aims to evaluate the performance of several Recurrent Neural Network (RNN) architectures, including Simple RNN, Gated Recurrent Units (GRU), and Long Short-Term Memory (LSTM), compared to classic algorithms such as Random Forest and XGBoost, in building classification models for early crash detection in the ASEAN-5 stock markets.

**Methods:** The study examines imbalanced data, which is expected due to the rarity of market crashes. It analyzes daily data from 2010 to 2023 across the major stock markets of the ASEAN-5 countries: Indonesia, Malaysia, Singapore, Thailand, and the Philippines. A market crash is the target variable when the primary stock price indices fall below the Value at Risk (VaR) thresholds of 5%, 2.5%, and 1%. Predictors include technical indicators from major local and global markets and commodity markets. The study incorporates 213 predictors with their respective lags (5, 10, 15, 22, 50, 200) and uses a time step of 7, expanding the total number of predictors to 1,491. The challenge of data imbalance is addressed with SMOTE-ENN. Model performance is evaluated using the false alarm rate, hit rate, balanced accuracy, and the precision-recall curve (PRC) score.

**Result:** The results indicate that all RNN-based architectures outperform Random Forest and XGBoost. Among the various RNN architectures, Simple RNN is the most superior, primarily due to its simple data characteristics and focus on short-term information.

**Novelty:** This study enhances and extends the range of phenomena observed in previous studies by incorporating variables such as different geographical zones and periods and methodological adjustments.

**Received** April 2024 / **Revised** June 2024 / **Accepted** June 2024

## INTRODUCTION

Early crash detection in financial markets is a crucial area of research with the potential to impact risk management and investment strategies significantly. Predicting sudden and substantial price drops, known as market crashes, is paramount for investors, financial institutions, and regulators. As financial markets become increasingly complex and interconnected, the need for accurate and timely crash prediction models has intensified [1], [2]. Crashes are rare, unexpected events like the 2007 subprime mortgage crisis or the 2020 COVID-19 pandemic. Early warning of these events can mitigate potential losses [3]. Artificial intelligence (AI) models and intense learning approaches have emerged as promising tools for addressing this challenge due to their ability to analyze vast amounts of data, identify patterns, and make predictions with increasing accuracy [4].

Deep learning models, particularly artificial neural networks (ANNs), mimic the human brain's functionality and serve as the cornerstone of the deep learning approach. Introduced in the 1990s, ANNs have experienced rapid development over the past two decades due to advancements in computing power, AI technology, and data availability [5]. Recurrent neural networks (RNNs), a notable ANN architecture, are well-suited for early crash detection, as they excel at predicting patterns in sequential data [6]. Early crash detection falls under the category of sequence classification, similar to DNA sequence classification and sentiment analysis [7].

---

[*]Corresponding author.

Email addresses: derideri@apps.ipb.ac.id (Siswara)[*], agusms@apps.ipb.ac.id. (Soleh),
aji_hw@apps.ipb.ac.id (Wigena)
DOI: 10.15294/sji.v11i3.4067

The architecture of RNNs has undergone several significant developments, with two of the most popular being Long Short-Term Memory (LSTM) and Gated Recurrent Units (GRU). LSTM can retain information for long periods and more effectively learn long-term dependencies in a dataset. Meanwhile, GRU is a simplified version of LSTM, which combines several gates in LSTM into one or two gates, making it more computationally efficient and often faster in training without sacrificing much capability in modeling long-term dependencies. GRU has proven effective in various applications, frequently delivering performance comparable to LSTM with a more compact structure. However, its effectiveness depends greatly on the specific case or application being addressed [8].

In the application to stock markets, the architecture of RNNs is frequently applied to predict stock price movements. RNNs are typically used to forecast future prices, like predicting the next day's stock prices or the trend for several days ahead. This application is reflected in studies by Li and Qian [9] and Jin et al. [10]. RNNs are also utilized to predict whether stock prices will go up or down, a binary trend prediction, as researched by Zhao et al. [11]. However, a significant research gap exists in predicting substantial and unusual price drops, known as market crashes, for early crash detection. This area still needs to be explored compared to modeling price movements.

Research on early crash detection in financial markets frequently utilizes machine and deep learning techniques. Bluwstein et al. [12] employed machine learning to predict recessions, while Tölö [13] used RNNs to detect financial crises, focusing on a broad range of macroeconomic factors. Their studies primarily explored the financial market rather than specifically targeting the stock market. Chatzis et al. [14] used neural networks and support vector machines with daily data from 39 countries for stock market-focused research, finding that deep neural networks significantly enhance classification accuracy. Meanwhile, Moser [15] examined RNN-LSTM models alongside classic machine learning, revealing that simpler models can be more effective than complex ones in various scenarios.

Additionally, Dichtl et al. [16] demonstrated that support vector machines are particularly effective in predicting stock market crashes in the Eurozone. Overall, findings suggest that while complex models can offer high classification accuracy, simpler machine learning models sometimes prove more effective in specific scenarios, demonstrating the need for a tailored approach depending on the particular data characteristics, architecture, or market specificity. Nonetheless, these studies indicate that there is still room for improvement in overall performance.

This research focuses on the early detection of market crashes in stock markets, defining a market crash as an anomaly or a significant price decline that falls outside an investor's risk tolerance. The study compares the performance of several RNN-based architectures—including Simple RNN, GRU, and LSTM—with classic machine learning algorithms such as Random Forest and XGBoost. This study enhances and extends the range of phenomena observed in previous research by incorporating variables like different geographical zones and periods and methodological adjustments. Technical indicators from both local and global financial markets are utilized as predictors. The stock market samples used in this research are from the ASEAN-5 zones. The occurrence of market crashes leads to imbalanced target data classes since a market crash is rare.

Consequently, this study also addresses the presence of imbalanced data classes, enabling the examination of RNNs and classic machine learning characteristics in such contexts. Philosophically, this study posits that for cases involving temporal or time series data, sequence-based models like RNNs are more appropriate than classic machine learning approaches [17]. Random Forest and XGBoost are employed to provide a comparative analysis, with XGBoost being particularly noted for its robust handling of imbalanced data classes [18]. This aspect makes it especially interesting to explore the comparative performance of these models.

**METHODS**
**Data and research design**
This study focuses on data from the five largest stock markets in the ASEAN-5 region, which includes Indonesia, Malaysia, the Philippines, Singapore, and Thailand, creating a total of five datasets. The data are publicly available from Yahoo Finance [19], making this research replicable. The data consist of daily stock prices, which are irregular due to stock market trading occurring only on weekdays, spanning from the year 2010 to 2023. The target variable, market crash, is determined based on the decrease in the primary stock price indices below specified Value at Risk (VaR) thresholds [14], [20], [21], which are 5%, 2.5%, and 1%. These thresholds reflect the level of risk that might be acceptable to investors. For instance, a 5% VaR threshold indicates that, based on the historical distribution of returns, the lowest 5% are considered a market crash (1: Positive), while the remainder is considered normal (0: Negative). Thus, the target variable is characterized as binary data. Similar scenarios are also applied for 2.5% and 1% VaR thresholds. Based on three imbalance class scenarios, there are 15 datasets (5 markets × 3 VaR scenarios). The illustration of the VaR threshold is shown in Figure 1.

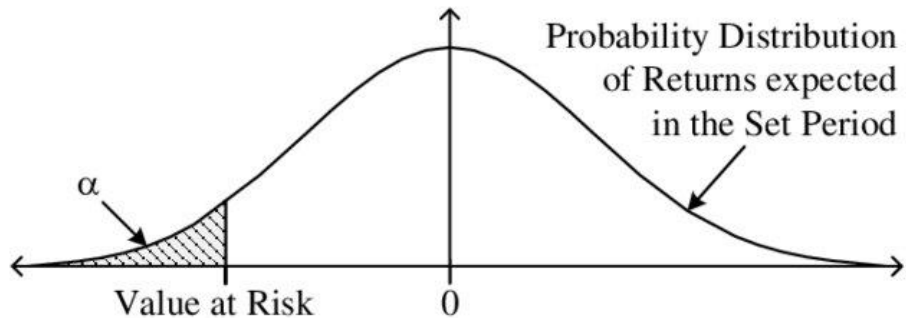

Figure 1. Illustration of the value at risk (VaR) threshold [22]

The predictors for the models include technical indicators such as return, moving average (MA), exponential moving average (EMA), the difference between opening and closing stock prices, relative strength index (RSI), and moving average convergence divergence (MACD). These indicators are derived from major local and global stock markets, currency exchange rates, and commodity markets. The global markets included in the predictors are the Dow Jones Industrial Average (DJI), NASDAQ, European markets (EURO), Japanese markets (JPAN), and the Green Index (FAN), while the considered commodities include crude oil, gold, and bonds. Currency exchange rates involve the value of each ASEAN-5 country's currency relative to the US dollar. MA and EMA are further subdivided into lags of the $5^{th}$, $10^{th}$, $15^{th}$, $20^{th}$, $22^{nd}$, $50^{th}$, and $200^{th}$ day, resulting in 213 predictors. The analysis uses time steps of 7, meaning each predictor has seven lags (from $X_{t-1}$ to $X_{t-7}$), leading to 1,491 predictors for each dataset. The current time point $X_t$ is excluded to focus the model on predicting the following day's outcomes, enabling market crash early detection one day in advance.

**Modelling**
This research utilizes deep learning models based on Recurrent Neural Networks (RNNs)—including Simple RNN, Long Short-Term Memory (LSTM), and Gated Recurrent Units (GRU) architectures—as well as ensemble machine learning models such as Random Forest and XGBoost. The Simple RNN represents the basic architecture of RNNs, where each hidden layer is interconnected. However, Simple RNNs often encounter issues with gradient vanishing. LSTM and GRU are sophisticated variants of RNNs designed to overcome this problem by incorporating gating mechanisms to regulate information flow. LSTMs utilize a memory cell and three types of gates (input, output, and forget) to preserve long-term information. At the same time, GRUs streamline this approach by merging several gates into a single unit. These details are depicted in Figure 2, which illustrates the various types of recurrent nodes used in RNNs. It showcases a non-gated recurrent node, exemplified by the Simple RNN, alongside gated recurrent nodes such as LSTM and GRU. Each node type is essential for processing sequential data, with gated nodes offering advanced capabilities for managing information flow and effectively retaining longer data sequences [23], [24].

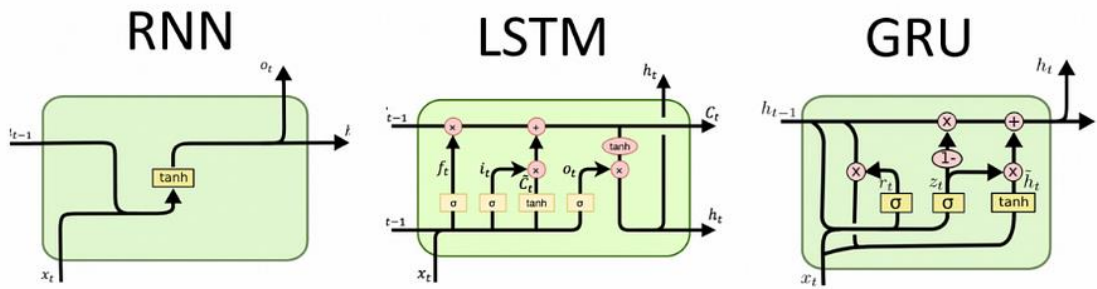

Figure 2. Recurrent nodes: gated and non-gated variants [25]

The study also employs Random Forest and Extreme Gradient Boosting (XGBoost) algorithms, critical players in ensemble machine learning. Random Forest improves model accuracy and mitigates overfitting by combining multiple decision trees, each trained on a random subset of the training data. Predictions are finalized either by averaging, in the case of regression, or by majority voting for classification tasks. This algorithm excels in handling datasets with numerous features and shows strong performance even with unbalanced data. Similarly, XGBoost is a robust boosting algorithm that sequentially combines predictions from multiple weak models to enhance overall prediction accuracy. Each subsequent model in the sequence focuses on addressing errors identified in the projections of its predecessors. XGBoost stands out due to its computational efficiency, capability to manage missing data effectively, and exceptional performance across a diverse range of machine learning applications, particularly those involving imbalanced datasets. These characteristics make Random Forest and XGBoost invaluable for tackling complex predictive problems where traditional single-model approaches might falter [26], [27].

**Evaluation metric and model architectures**
Various models were implemented across five ASEAN-5 stock market datasets. The best-performing model was selected based on four primary metrics: precision-recall curve (PRC), balanced accuracy, false alarm rate, and hit rate [14]. The PRC is critical for assessing the trade-off between precision and recall, particularly for the minority class, which is highly relevant for imbalanced datasets. Balanced accuracy offers a fair assessment of model performance across both classes. The false alarm rate measures the frequency of incorrect market crash predictions. In the study, the false alarm rate is inverted to facilitate interpretation. The hit rate assesses the model's ability to detect market crashes accurately. The formulas are shown in Table 1.

Table 1. Metrics evaluation

| No. | Metrics | Formula | |
|---|---|---|---|
| 1 | Inverted False Alarm Rate | $1 - \frac{FP}{FP+TN}$ | (1) |
| 2 | Hit Rate | $\frac{TP}{TP+FN}$ | (2) |
| 3 | Balanced Accuracy | $\frac{1}{2}\left(\frac{TP}{TP+FN} + \frac{TN}{TN+FP}\right)$ | (3) |
| 4 | Precision-Recall Curve (PRC) Scores.* | $Recall = \frac{TP}{TP+FN}$ | (4) |
| | | $Precision = \frac{TP}{TP+FP}$ | (5) |

*The area under the PRC (AUC-PRC) scores provides a single measure of overall model performance regarding the trade-off between precision and recall.

Where true positives (TP) is the number of correct positive predictions, false positives (FP) is the number of incorrect positive predictions, false negatives (FN) is the number of incorrect negative predictions, and true negatives (TN) is the number of correct negative predictions. Where TP is the number of correct positive predictions, FP is the number of incorrect positive predictions, FN is the number of incorrect negative predictions, and TN is the number of correct negative predictions.

Determining the best hyperparameters for each model was meticulously conducted through a grid search process, which involved iterating over a predefined range of values for each model's architecture and algorithm-specific settings. The configurations for the RNN models—Simple RNN, LSTM, and GRU—included a dense output layer of one unit, the Adam optimizer, and a time step setting of seven. Activation functions were set to ReLU for internal layers and Sigmoid for the output layer. The input shape was consistent across models, defined by the number of time steps and features. Neuron counts varied across three scales: 32, 64, and 128, while the architectures were tested with one or two layers. Learning rates considered were 0.001, 0.01, and 0.1, with training regimes allowing up to 50 epochs and employing early stopping to prevent overfitting [28]. The early stopping configuration is set to monitor the validation precision-recall curve. Training stops if no improvement is seen after ten epochs and the best model weights are restored. The models incorporate L1 and L2 regularization to mitigate potential overfitting by constraining the size of the model coefficients through the kernel, recurrent, and bias regularization [29].

Different configurations were tested for the machine learning models, Random Forest and XGBoost. Random Forest was examined with varying numbers of estimators (100, 200, 300) and maximum depth levels (10, 20, 30). Similarly, XGBoost configurations varied across estimators (100, 200, 300), learning rates (0.01, 0.1, 0.2), and maximum depths (3, 4, 5). These hyperparameter grids were chosen to explore the trade-offs between model complexity, performance, and overfitting, providing a comprehensive foundation for selecting the optimal model configurations. The details are shown in Table 2.

Table 2. Architecture and grid hyperparameters for each model

| Architectures / Algorithms | Hyperparameter |
|---|---|
| Simple RNN, LSTM, and GRU | Dense (Output Layer): 1 unit |
| | Optimizer: Adam |
| | Time Steps: 7 |
| | Activation Function: ReLU for layers, Sigmoid for the output layer |
| | Input Shape: (Time Steps, Number of Features) |
| | Number of Neurons: 32, 64, 128 |
| | Number of Layers: 1, 2 (applied to Simple RNN, LSTM, GRU) |
| | Learning Rate: 0.001, 0.01, 0.1 |
| | Epochs: 50, with early stopping enabled |
| | Kernel Regularization: L1 = $1\times10^{-5}$, L2 = $1\times10^{-4}$ (applied to kernel, recurrent, and bias weights) |
| Random Forest | N Estimators: 100, 200, 300 |
| | Max Depth: 10, 20, 30 |
| XGBoost | N Estimators: 100, 200, 300 |
| | Learning Rate: 0.01, 0.1, 0.2 |
| | Max Depth: 3, 4, 5 |

Each analysis in this research began with the model implementation without addressing the issue of imbalanced data expressed as a baseline. This initial step is influential for understanding how the models perform under the dataset's natural conditions. Subsequently, to overcome the challenges caused by the imbalanced data condition, a specific treatment was applied utilizing the Synthetic Minority Over-sampling Technique (SMOTE) in conjunction with Edited Nearest Neighbors (ENN). This method combines an oversampling technique, which adds samples to the minority class, with a cleaning technique to remove ambiguous or overlapping samples. Hence, further analyses were conducted using imbalanced data handling to observe any improvements or changes in model performance based on the predefined metrics. The use of SMOTE-ENN has been demonstrated to be effective, as evidenced by the research of Mukhlashin et al. [30].

**The stages of analysis**
The stages of analysis in this research are visualized in Figure 3 and described as follows:

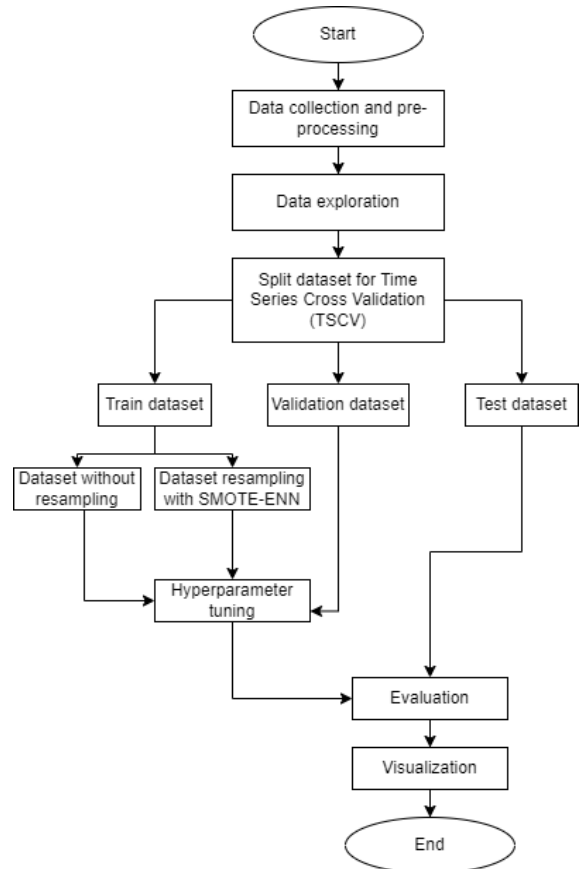

Figure 3. A flowchart of the analysis stages

1. Data collection and data pre-processing
   a. Each of the datasets is sourced from Yahoo Finance.
   b. Form predictor variables consisting of technical indicators from each dataset [31].
   c. Establish target variables with three scenarios based on Value at Risk (VaR) return thresholds: 5% (moderate), 2.5% (high), and 1% (extreme) and labeled as "Crash" if the return is below the specified threshold.
   d. The imputation process is used to complete missing data caused by operational time differences between local and global markets using K-Nearest Neighbors (KNN) [32]. This imputation generates at most 20% of the data because variables with a missing data proportion exceeding 20% have been previously eliminated [33].
2. Data exploration
3. Data is split into training, validation, and testing data to perform the Time Series Cross-Validation (TSCV) [34]. Training and validation data are combined to find the optimal hyperparameters, while training and testing data are combined to evaluate final performance. The data partition details are shown in Table 3.

Table 3. Data partition details

| K | Data training period | Data validation period |
|---|---|---|
| 1 | 01-01-2010----31-12-2011 | 01-01-2012----31-12-2013 |
| 2 | 01-01-2010----31-12-2013 | 01-01-2014----31-12-2015 |
| 3 | 01-01-2010----31-12-2015 | 01-01-2016----31-12-2019 |
|   |   | Data testing period |
| 4 | 01-01-2010----31-12-2019 | 01-01-2020----31-12-2023 |

4. Analysis of the ASEAN-5 datasets with Simple RNN, LSTM, and GRU architectures, as well as Random Forest and XGBoost algorithms through the TSCV process, so that results evaluation metrics for each architecture/algorithm.
5. Visualization of early crash detection with the foremost architecture/algorithm.

**Implementation of methods**
Simple RNN, LSTM, and GRU models were implemented using TensorFlow, a widely used deep learning framework known for its efficient handling of sequential data and complex neural network architectures [35], [36]. The specific architectures are detailed in Table 3. The Random Forest implementation was carried out using the Scikit-learn library, a popular Python machine-learning toolkit renowned for its ease of use and robustness in implementing Random Forest algorithms [37]. XGBoost, another widely respected library, was also utilized for its capabilities [38]. All programming was done in Python, with experiments conducted on the VS Code IDE. This setup leveraged the flexibility of local computing for efficient analysis and model training. The computational tools used included a 12th Gen Intel(R) Core(TM) i5-1240P processor with 16.0 GB of DDR5 RAM, running on a Windows 11 Home edition system and equipped with a Generation 4 SSD Storage.

**RESULTS AND DISCUSSIONS**
**Data exploration**
Value at Risk (VaR) scenarios establish different return thresholds for defining crashes in each country's dataset, as shown in Table 4. This finding reveals varying patterns in the return distribution of each stock market. These differences may arise from investors' investment patterns or policies implemented by the stock market authorities in each country.

Table 4. Datasets' crash return thresholds are based on their respective Value at Risk (VaR) scenarios

| Datasets | VaR 5% | VaR 2.5% | VaR 1% |
| --- | --- | --- | --- |
| Indonesia | -1.62% | -2.24% | -3.13% |
| Malaysia | -1.06% | -1.36% | -1.85% |
| Philippines | -1.70% | -2.25% | -3.02% |
| Singapore | -1.28% | -1.58% | -2.22% |
| Thailand | -1.47% | -1.95% | -2.78% |

Various instances of stock market crashes have occurred in the ASEAN-5 zone, with patterns of crash occurrences that are relatively similar, as shown in Figure 4. Exploring the points of crash occurrences in each market forms the basis for data partitioning for Time Series Cross-Validation (TSCV) and also the determination of predictors. The data partitioning has irregular intervals, as there are years when no crashes occurred at all in some markets. The global financial crisis is the leading cause of crashes in the ASEAN-5 stock markets. These markets are relatively easily shaken simultaneously by global shocks, considering that the stock markets in this zone are still developing. Some global financial crises that align with the occurrences of crashes in the ASEAN-5 stock markets include the Chinese Stock Market Crisis (2015–2016), the US Market Sell-Off (2015–2016), and the COVID-19 Pandemic (2020).

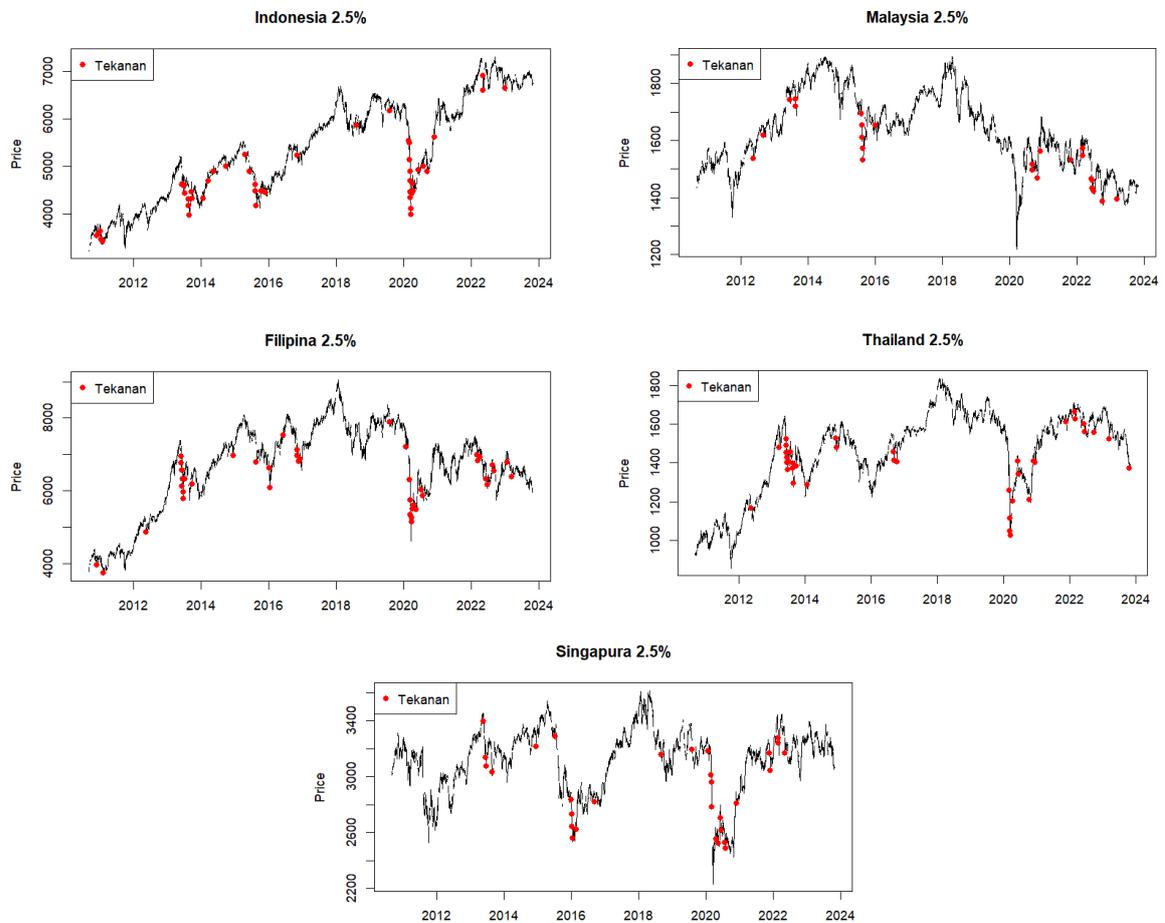

Figure 4. The crashes that occurred in the ASEAN-5 stock markets under the 2.5% VaR scenario

**Evaluation**

This section presents the performance of each model across various scenarios. To elaborate, we executed each model on its dataset ten times, incorporating regularization strategies to mitigate overfitting risks. Implementing these strategies was essential to fortify the models' learning processes, making them more resilient and less prone to assimilating noise or irrelevant patterns from the training data. The initial analysis was conducted using various algorithms and architectures without considering the class imbalance of the data, which means that no resampling process was performed on the dataset to handle the imbalance issue. This initial analysis acts as a baseline. The baseline evaluation revealed that the RNN-based architecture failed to predict stock market crashes (hit rate = 0). Only Random Forest succeeds in the 2.5% VaR scenario despite the trade-off of a low false alarm rate.

Consequently, the baseline results reveal the need for imbalanced data handling strategies to improve the quality of predictive models. Subsequent analysis integrating the SMOTE-ENN approach shows substantial performance improvements. As illustrated in Figure 5, the distribution of performance values across the ASEAN-5 dataset with a 1% VaR crash scenario significantly improves the evaluation metrics after applying the SMOTE-ENN technique.

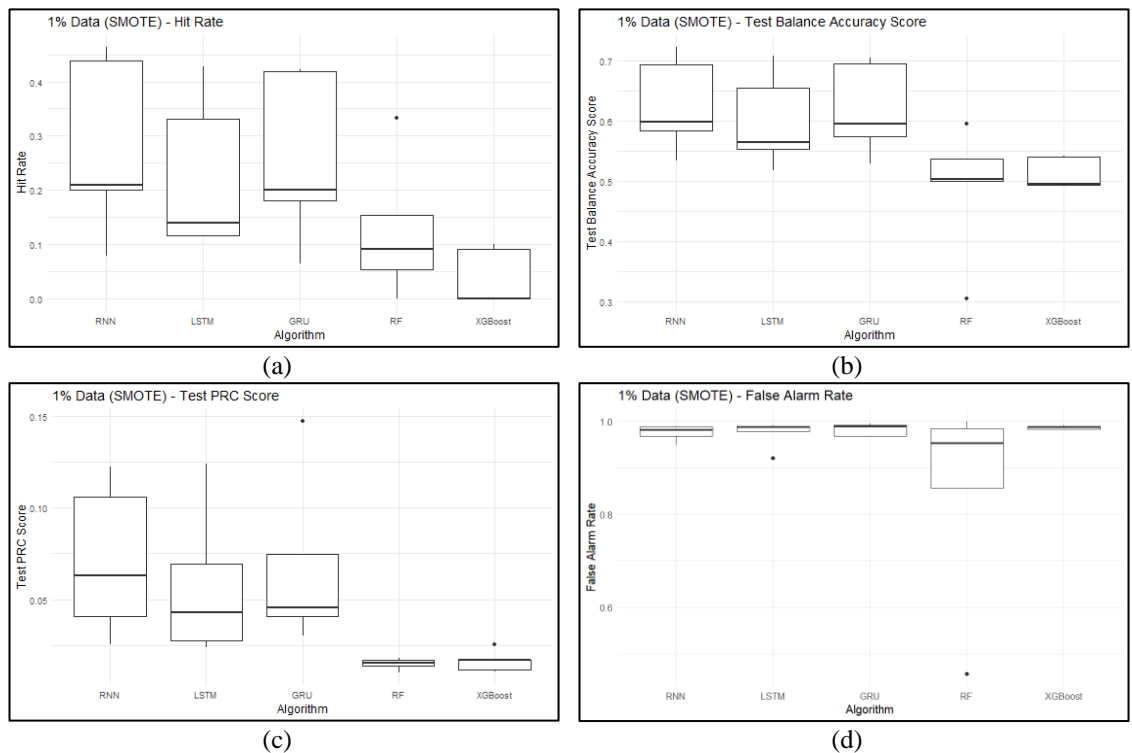

(a)                                                    (b)

(c)                                                  (d)

Figure 5. Performance evaluation on datasets with VaR 1% ASEAN-5 based on metrics
(a) Hit Rate, (b) Balance accuracy score, (c) PRC score, and (d) False alarm rate

The RNN-based architectures (RNN, LSTM, and GRU) demonstrated higher hit rate, balanced accuracy, and PRC scores compared to Random Forest and XGBoost. This indicates their effectiveness in predicting market crashes one day ahead within the ASEAN-5 dataset, although with marginally lower false alarm rates relative to XGBoost. Table 5 concisely encapsulates the performance comparison by presenting average values for these metrics, which provides an integrated perspective of the relative efficacy of each algorithm/architecture.

Table 5. Average values of each algorithm's/architecture's performances on every dataset

| Algorithms/architectures | False Alarm Rate | Hit Rate | Balance Accuracy | PRC Score |
|---|---|---|---|---|
| **Dataset 1%** | | | | |
| RNN | 0.975 | **0.278** | **0.626** | **0.072** |
| LSTM | 0.973 | 0.226 | 0.600 | 0.058 |
| GRU | 0.982 | 0.257 | 0.620 | 0.068 |
| RF | 0.850 | 0.126 | 0.488 | 0.015 |
| XGBoost | **0.987** | 0.038 | 0.512 | 0.016 |
| **Dataset 2.5%** | | | | |
| RNN | 0.853 | 0.362 | **0.608** | 0.066 |
| LSTM | 0.834 | **0.366** | 0.600 | **0.067** |
| GRU | 0.846 | 0.343 | 0.594 | **0.067** |
| RF | **0.988** | 0.008 | 0.498 | 0.028 |
| XGBoost | 0.971 | 0.108 | 0.539 | 0.040 |
| **Dataset 5%** | | | | |
| RNN | 0.885 | 0.266 | **0.575** | **0.074** |
| LSTM | 0.867 | **0.276** | 0.571 | 0.067 |
| GRU | 0.902 | 0.260 | 0.581 | 0.072 |
| RF | **0.982** | 0.064 | 0.523 | 0.059 |
| XGBoost | 0.943 | 0.149 | 0.546 | 0.062 |

*Thickened values show the leading evaluation values among all algorithms/architectures

The performance analysis generally shows that models based on recurrent neural networks (RNN, LSTM, and GRU) surpass Random Forest and XGBoost models in hit rate, balanced accuracy, and PRC score. These results are consistent across datasets with varying Value at Risk (VaR) crash thresholds, indicating a superior capability of RNN-based models in identifying market crashes. While XGBoost has the highest false alarm rate in the 1% VaR dataset, Random Forest shows the highest false alarm rate in the 2.5% and 5% VaR datasets, indicating its better tendency to reduce false crashes. The false alarm rate of RNN-Based is considered to be within acceptable limits. Eventually, the accuracy in the early detection of market crashes (hit rate) is prioritized as a metric due to its importance in minimizing potential investment losses [14].

The RNN-based architecture is superior to Random Forest and XGBoost in this study because it is well-suited for modeling time dependencies. RNN-based models are more effective than tree-based models in capturing time dynamics [39]. RNNs can handle variable-length sequences in time series data that often contain irregular sampling intervals [40]. This flexibility is not inherently available in models like Random Forest or XGBoost, which typically rely on fixed-length feature vectors [41].

After that, in the case of comparing the performance among the RNN-based architectures, Simple RNN demonstrates better performance than its developments, LSTM and GRU. LSTM slightly outperforms in hit rate for datasets with 2.5% and 5% VaR, but Simple RNN maintains its superiority in balanced accuracy. This finding suggests that the datasets used in the research might not be overly complex or not require extensive processing of long-term historical information, aligning with the basic capabilities of Simple RNN that prioritize short-term memory [42].

Further analysis of each country within ASEAN-5 shows that RNN-based performance is generally superior to Random Forest and XGBoost algorithms, although there is a slight exception in the Thailand dataset (Figure 6). On the Thailand dataset with 1% VaR, Random Forest recorded the highest hit rate and balanced accuracy, beating RNN-based. These results suggest the potential to develop better architectures or apply different architectures depending on the study and the characteristics of stock market data in ASEAN-5, which may differ.

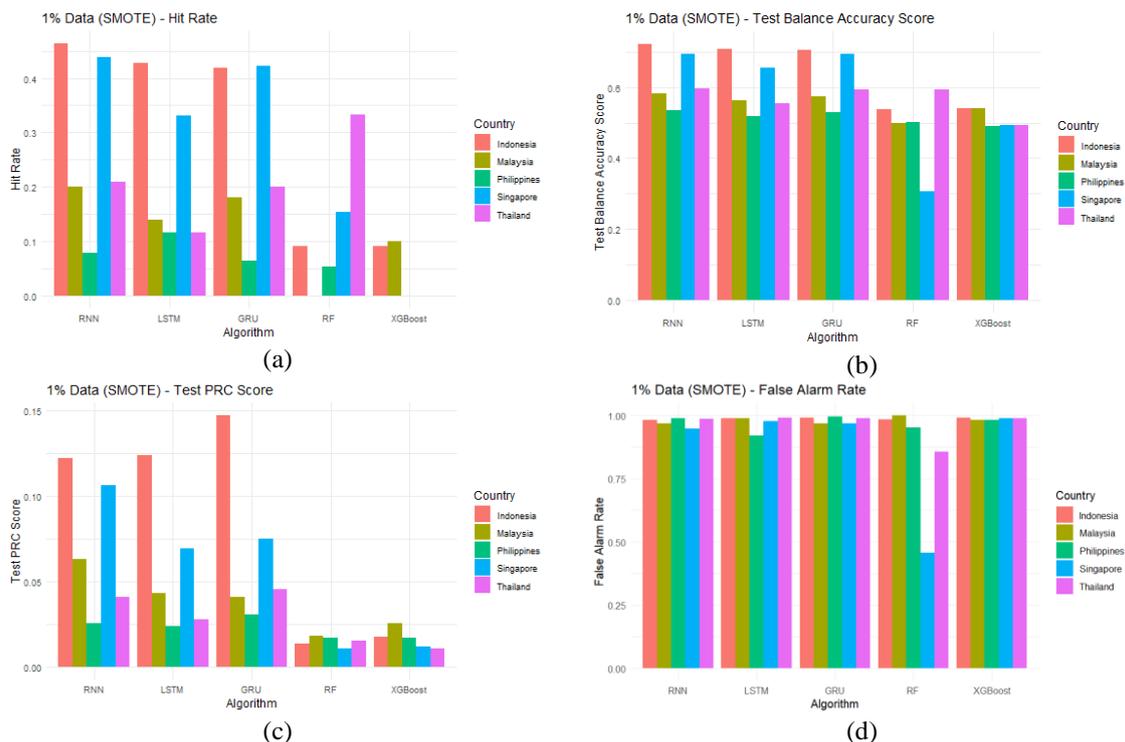

Figure 6. Performance evaluation on datasets with VaR 1% scenario of each country with metrics
(a) Hit rate, (b) Balance accuracy score, (c) PRC score, and (d) False alarm rate

Overall, the performance of the hit rate metric in this study is in the interval of 0%–64%, with an average of 21%. This value is undoubtedly still relatively low, and better model and architecture development needs to be developed. However, the value is still at a consistent interval compared to previous research. Research by Chatzis et al. [14] has a hit rate value interval of 38%–59%, Moser's [15] has a hit rate value interval of 45%–71%, and research by Dichtl et al. [16] has an interval of 9%–50% hit rate value. The hit rate value in this study closely aligns with that reported by Dichtl et al. [16], as both analyses encountered datasets from the same period of the financial crisis triggered by COVID-19, which introduced significant anomalies in the stock market.

Developing an architecture focusing on RNN-based types, such as bidirectional and stacked techniques, is essential to enhance performance [43]. Another effort that can be made is to extend the time steps so that the information stored can be more comprehensive in evaluating the performance of the LSTM and GRU models [44]. Finally, for each case and each country's different data characteristics, it is necessary to adjust the basic architecture and model individually. Future researchers need to capture continuous crash predictions (>1 day).

**Visualization**
The market crash early detection model becomes more helpful when the crash probability visualization is combined with the best algorithm or architecture [45]. Figure 7 displays the visualization of the early detection model using RNN for the Indonesian dataset with a 1% VaR crash over 2020–2023. In this case, the RNN demonstrated the ability to detect crashes one day in advance with a hit rate that did not exceed 50% of the total crash incidents. However, the Simple RNN was significantly successful in identifying market crashes during the crisis caused by the COVID-19 pandemic in March 2020, with few false alarms. False alarms in crisis periods can help investors raise alertness. The RNN experienced several failures in detecting crashes outside the crisis period, indicating limitations in long-term post-crisis predictions. It cannot effectively foresee those distinctive patterns that have yet appeared previously [46].

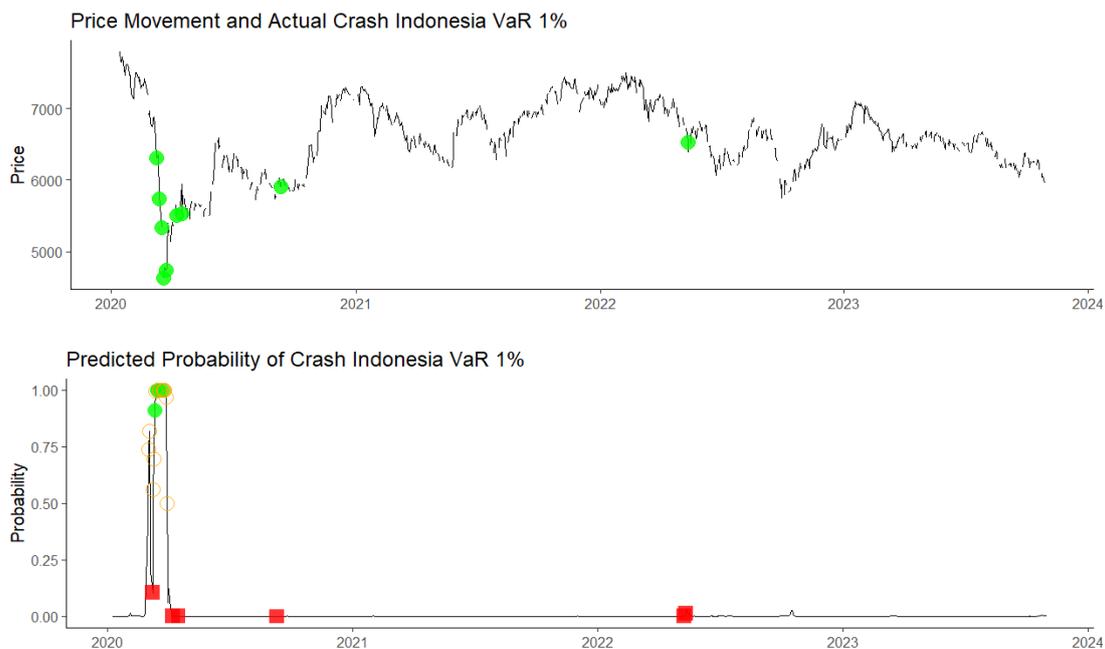

Figure 7. The early crash detection models for the Indonesian dataset with VaR 1% using Simple RNN architectures during 2020–2023

**CONCLUSION**
The primary objective of this study was to assess the performance of various recurrent neural network (RNN) architectures, including Simple RNN, Gated Recurrent Units (GRU), and Long Short-Term Memory (LSTM), in comparison to classical machine learning algorithms such as Random Forest and XGBoost. This assessment focused on their application in building classification models for early crash detection in the ASEAN-5 stock markets. The study has conclusively demonstrated that RNN-based architectures, particularly the Simple RNN, provide superior performance in the context of early crash detection. The Simple RNN was notably effective due to its ability to handle the less complex data characteristics of the ASEAN-5 stock markets and its focus on short-term information. The model achieved commendable accuracy metrics, including a balanced accuracy of 64%, and demonstrated a strong capacity to detect significant market downturns, even under the challenging conditions of imbalanced data. This research substantially contributes to the field by extending the understanding of how different RNN architectures can be leveraged to enhance early crash detection models. It introduces novel methodological adjustments and incorporates various predictors across different geographical zones and periods, enriching the financial analytics landscape in emerging markets. The findings of this study underscore the potential of RNN-based models in stock market applications and provide a robust foundation for future research aimed at refining these models for even greater accuracy and applicability in modeling early crash detection.